\documentclass{article}
\usepackage[left=0.5in,right=0.5in,bottom=0.75in,top=0.75in]{geometry}

\usepackage{amsmath}
\usepackage{amssymb}
\usepackage{amsthm}
\usepackage{graphicx}
\graphicspath{{fig/}}
\usepackage{natbib}
\usepackage{hyperref}
\hypersetup{colorlinks=true, citecolor=black, linkcolor={}, urlcolor=cyan}
\usepackage[font = small]{caption}
\usepackage[shortlabels]{enumitem}
\usepackage{booktabs}
\usepackage{appendix}
\usepackage{tikz}
\usetikzlibrary{decorations.pathreplacing}

\providecommand{\A}{\mathbf{A}}

\providecommand{\C}{\mathbf{C}}

\providecommand{\I}{\mathbf{I}}

\providecommand{\K}{\mathbf{K}}

\renewcommand{\S}{\mathbf{S}}

\providecommand{\X}{\mathbf{X}}

\providecommand{\Z}{\mathbf{Z}}

\renewcommand{\l}{\mathbf{l}}

\newcommand{\x}{\mathbf{x}}
\newcommand{\y}{\mathbf{y}}
\newcommand{\z}{\mathbf{z}}
\let\origc\c
\DeclareRobustCommand\c{\ifmmode\mathbf{c}\else\expandafter\origc\fi}
\let\origd\d
\DeclareRobustCommand\d{\ifmmode\mathbf{d}\else\expandafter\origd\fi}
\let\origu\u
\DeclareRobustCommand\u{\ifmmode\mathbf{u}\else\expandafter\origu\fi}
\let\origd\v
\DeclareRobustCommand\v{\ifmmode\mathbf{v}\else\expandafter\origv\fi}

\providecommand{\byy}{\tilde{\mathbf{y}}}

\providecommand{\XX}{\widetilde{\mathbf{X}}}

\providecommand{\ba}{\boldsymbol{\alpha}}

\providecommand{\bb}{\boldsymbol{\beta}}

\providecommand{\bbh}{\widehat{\boldsymbol{\beta}}}
\providecommand{\bd}{\boldsymbol{\delta}}

\providecommand{\bvep}{\boldsymbol{\varepsilon}}

\providecommand{\bg}{\boldsymbol{\gamma}}

\providecommand{\lam}{\lambda}

\providecommand{\bS}{\boldsymbol{\Sigma}}
\providecommand{\bSh}{\hat{\boldsymbol{\Sigma}}}

\providecommand{\btau}{\boldsymbol{\tau}}

\providecommand{\Ex}{\mathbb{E}}

\providecommand{\Var}{\mathbb{V}}

\providecommand{\Norm}{\textrm{N}}

\providecommand{\Tr}{^{\scriptscriptstyle\top}}

\renewcommand{\l}{\ell}

\usepackage[scr=boondox]{mathalpha}

\makeatletter
\@ifpackagelater{mathalpha}{2021/01/01}{%

}{%

}
\makeatother

\providecommand{\one}{\mathbf{1}}
\providecommand{\zero}{\mathbf{0}}

\providecommand{\real}{\mathbb{R}}

\providecommand{\given}{\,\vert\,}
\providecommand{\abs}[1]{\left\lvert#1\right\rvert}

\providecommand{\norm}[1]{\lVert#1\rVert}

\providecommand{\as}[1]{\begin{align*}#1\end{align*}}
\providecommand{\als}[2]{\begin{align}\label{#1}\begin{split}#2\end{split}\end{align}}

\renewcommand{\abstract}[1]{
 \centerline{
 \begin{minipage}{0.7\linewidth}
 \hrule
 \vskip 0.1in
  \begin{center}
    {\bf Abstract}
  \end{center}
  #1
 \vskip 0.1in
 \hrule
 \end{minipage}}
 \vskip 0.3in}

\title{Penalized mixed models to adjust for batch effects and unobserved confounding in high dimensional regression}
\author{
  Yujing Lu\\Department of Biostatistics\\University of Iowa
  \and
  Patrick Breheny\\Department of Biostatistics\\University of Iowa
}
\date{\today}

\begin{document}

\maketitle

\abstract{Confounding can lead to spurious associations. Typically, one must observe confounders in order to adjust for them, but in high-dimensional settings, recent research has shown that it becomes possible to adjust even for unobserved confounders. The methods for carrying out these adjustments, however, have not been thoroughly investigated. In this study, we derive a confounding framework in which the signal, bias, and variability can be cleanly partitioned and quantified, thereby enabling simulations in which one varies the bias-to-signal ratio while holding the signal-to-noise ratio fixed. Using this construction, we demonstrate the impact of the amount and complexity of unobserved confounding on the performance of competing methods, including the LASSO, principal components LASSO (PC-LASSO), and penalized linear mixed models (PLMMs). We identify scenarios in which each method outperforms the others and find that overall, PLMM is the most robust approach.}


\section{Introduction}

Confounding leads to spurious associations and estimation bias if not properly adjusted for. Traditionally, adjusting for confounders relies on including them in a regression model. To do so, however, the confounders must be observed. Recently, research has shown that in high-dimensional settings, it is possible to adjust for confounders even when they are unobserved. \citet{Wang2019} demonstrated how high-dimensional data could be used to estimate confounding structures and proposed a three-step deconfounding method using factor analysis techniques. A similar phenomenon arises in the context of batch effects for high-throughput data \citep{Leek2010}, where approaches have been developed to remove batch effects even when batches are unknown \citep{Leek2008,Gagnon-Bartsch2012}.

In penalized regression, principal components analysis (PCA) is the most widely used approach to adjusting for confounding effects. This is accomplished by including the top principal components (PCs) as unpenalized variables in the model. One caveat of PCA is that it requires choosing the number of PCs, which is not straightforward and may lead to inefficient adjustment. More recently, preconditioning techniques have been proposed, providing a new paradigm for handling confounding effects. The Puffer transformation \citep{Jia2015} and Lava estimator \citep{Chernozhukov2017} were among the first preconditioning approaches proposed, although their focus was on variable selection as opposed to confounding correction. \citet{Cevid2020} showed that these approaches, along with the Trim transformation, belong to a family of spectral transformations and that all of these approaches could be used to deal with hidden confounding structures.

Mixed models, despite being widely used in genetics, have not received much attention outside that context. The potential of mixed models in dealing with confounding effects outside of genome-wide association studies has not been investigated thoroughly.

This paper develops a rigorous framework for investigating unobserved confounding in high-dimensional regression that allows us to decompose the unobserved confounding effect into two components, bias and noise. Using this framework enables us to carry out simulations in which we vary the bias-to-signal ratio while holding the signal-to-noise ratio fixed, or vary the structure of the confounding while fixing the bias-to-signal and signal-to-noise ratios. These simulations highlight scenarios in which PC-LASSO outperforms PLMM and vice versa, while also demonstrating the ability of both methods to outperform the LASSO when observed confounders are present.

In Section~\ref{Sec:methods}, we describe the three methods under comparison. In Section~\ref{Sec:decomposition}, we formulate the framework described above, which we refer to as the linear confounding model, and illustrate how it can be decomposed into bias and noise. In the remaining sections, we compare the aforementioned methods using both simulated and real data.


\section{Methods} \label{Sec:methods}

In this manuscript, we consider the following setting:
\as{\y = \X\bb + \Z\bg + \bvep,}
where $\bvep \sim \Norm(\zero, \sigma_e^2\I_n)$ and the outcome $\y\in\real^n$ depends on both an observable $n\times p$ matrix of features $\X$ and an unobservable $n\times q$ matrix of confounders $\Z$, with effects $\bb \in \real^p$ and $\bg \in \real^q$, respectively. We refer to features with nonzero effects as signals. Throughout, we assume without loss of generality that $\X$ has been standardized such that $\sum_i x_{ij} = 0$ and $\sum_i x_{ij}^2/n = 1$ for all $j \in 1, ..., p$.

Since $\Z$ is not observed, however, we cannot directly include it in the model. Therefore, we consider various models in which the mean structure is misspecified as 
$$ \Ex \y = \X\bb. $$
Below, we introduce a standard LASSO (which does not address unobserved confounding) as well as two methods, PC-LASSO and PLMM, that attempt to adjust for unobserved confounding. 

\subsection{LASSO} 

We begin by giving a brief review of the LASSO \citep[Least Absolute Shrinkage and Selection Operator;][]{Tibshirani1996}. Unlike stepwise variable selection, which retains or discards covariates without shrinking coefficients, or ridge regression, which shrinks coefficients towards zero but does not perform variable selection, LASSO achieves both simultaneously. 

LASSO obtains its coefficient estimates by minimizing the objective function 
\as{
    Q(\bb\given\X, \y) = \frac{1}{2n}\norm{\y-\X\bb}_2^2 + \lambda\norm{\bb}_1,
}
where $\norm{\bb}_1 = \sum_{j=1}^p\abs{\beta_j}$ is the $\l_1$ norm and $\norm{\x}_2=\sqrt{\sum_{i=1}^n x_i^2}$ is the $\l_2$ norm. The tuning parameter $\lambda$ controls the trade-off between goodness-of-fit and shrinkage and is typically selected using cross-validation.

LASSO's ability to perform both variable selection and coefficient shrinkage allows it to retain both interpretability and prediction accuracy in high-dimensional settings where the number of covariates $p$ may be larger than the number of observations $n$. LASSO can be extended to handle other types of outcomes, such as counts or binary data, through appropriate generalized linear models. 

However, LASSO does not account for unobserved confounding. Confounding can affect the performance of LASSO in many ways, potentially leading to the selection of spurious features, bias in the estimation of $\bb$, and reduced prediction accuracy.

\subsection{PC-LASSO} 

One method to adjust for an unobserved $\Z\bg$ is to include principal components (PCs) derived from the observed features $\X$. The motivation is that unobserved confounding likely affects many features, and therefore the unobserved confounding may be captured by the leading PCs. Including them in the model will therefore adjust for the effect of confounding and allow for better estimation of $\bb$.

Let $\C$ denote an $n\times k$ matrix containing the first $k$ PCs. Then by incorporating $\C$ into the model, PC-LASSO estimates the coefficients by minimizing the objective function
\as{Q(\bb, \ba\given\X, \y) = \frac{1}{2n}\norm{\y-\C\ba-\X\bb}_2^2 + \lambda\norm{\bb}_1,}
where $\ba \in \real^k$ is the coefficient vector corresponding to the $k$ leading PCs. Note that $\ba$ is not penalized.

\subsection{PLMM} 

Like PC-LASSO, PLMM is motivated by the idea that unobserved confounding likely affects many features. However, whereas PC-LASSO adjusts for unobserved confounders through the mean structure, PLMM carries out its adjustment through the variance. Specifically, PLMM treats the unobserved confounding as a random effect with mean zero and $\Var(\Z\bg) = \sigma_s^2\K$. 

Under PLMM, therefore, the outcome follows the distribution $\y \sim \Norm(\X\bb, \bS)$, where $\bS=\sigma_s^2\K + \sigma_e^2\I_n$. Here, $\sigma_s^2$ controls the magnitude of the structured variance due to the confounding. Similar to generalized least squares models, we can transform the model to obtain a pseudo-response with independent errors:
\als{Eq:pre-multiply plmm}{
  \bS^{-1/2}\y \sim \Norm(\bS^{-1/2}\X\bb, \I_n). 
}
Letting $\byy = \bS^{-1/2}\y$ and $\XX = \bS^{-1/2}\X$ denote the pseudo-response and rotated features, respectively, the objective function for PLMM is then equivalent to that of the LASSO, and existing algorithms can be used to solve for $\bb$: 
$$ Q(\bb\given\XX, \byy) = \frac{1}{2n}\norm{\byy - \XX\bb}_2^2 + \lambda\norm{\bb}_1. $$


Estimating $\bS$ requires estimates of $\K$ as well as the variance components, $\sigma_s^2$ and $\sigma_e^2$. The matrix $\K$ is estimated by
$$ \hat{\K} = \frac{1}{p} \sum_j \x_j \x_j \Tr; $$
with standardized features, this estimates the $n \times n$ correlation structure among the instances. The variance components are typically estimated using the null model \citep{Lippert2011}. It is possible to estimate $\bS$ in other ways, but this is the most common approach and what is implemented in the software package used in later sections.

The correlation between observations also affects prediction. Suppose $\X_{\text{old}}$ and $\y_{\text{old}}$ are used for fitting the model, and $\X_{\text{new}}$ is the data used for predicting $\y_{\text{new}}$. Let $\hat{\bb}$ denote the estimated coefficients using $\X_{\text{old}}$. We can aggregate $\X_{\text{old}}$ and $\X_{\text{new}}$ to estimate $\bSh$, which consists of four blocks: the top-left block $\bSh_{11}$ is the estimated covariance matrix of $\X_{\text{old}}$; the bottom-right block $\bSh_{22}$ is the estimated covariance matrix of $\X_{\text{new}}$; off-diagonal blocks $\bSh_{12}$ and $\bSh_{21}$ represent the estimated covariance between $\X_{\text{old}}$ and $\X_{\text{new}}$. Based on the conditional distribution of the multivariate normal distribution, the best linear unbiased prediction (BLUP) of $\y_{\text{new}}\given\y_{\text{old}}$ is 
\as{
    \y_{\text{new}}\given\y_{\text{old}}=\X_{\text{new}}\bbh + \bSh_{21}\bSh_{11}^{-1}(\y_{\text{old}} - \X_{\text{old}}\bbh).
}
In cross-validation (CV), $\X_{\text{new}}$ corresponds to the data in the testing fold, and $\X_{\text{train}}$ consists of all other folds used as training data. $\bSh_{21}$ can be obtained by computing the covariance matrix between $\X_{\text{test}}$ and $\X_{\text{train}}$. And $\bSh_{11}$ is computed as the variance-covariance matrix of $\X_{\text{train}}$. 

\section{Decomposition of Unobserved Confounding} \label{Sec:decomposition}

Our goal in this paper is to examine how unobserved confounding affects LASSO estimates, to determine whether PC-LASSO and PLMM can correct for this confounding, and to investigate whether certain types of confounding favor one method over the other. However, confounding can affect the model in many ways: it can introduce bias, it can increase noise, the bias can be away from zero or towards zero, and it can be mistaken for signal, which affects selection of the regularization parameter $\lam$. Careful design is necessary to isolate effects such as increasing the amount of confounding, as misleading artifacts often arise when multiple aspects of the data are changing at the same time. In this section, we derive the mathematical model and decompositions that allow us to isolate aspects of unobserved confounding and derive useful quantities such as the bias to noise ratio that we utilize in the simulations that follow.

\subsection{Implications of Model Misspecification} \label{Sec:misspecification}

We begin by working out the implications of model misspecification on the estimation of $\bb$. We consider the features $\x = (x_1, ..., x_p)$, the confounders $\z = (z_1, ..., z_q)$, and the outcome $y$ as random variables. We further assume that ($\x, y, \z)$ has a joint distribution such that each dataset can be seen as multivariate realizations from this distribution. Without assuming any model, $y$ can be written as $y=\mu(\x, \z) + \epsilon$, where $\mu(\x, \z)$ denotes the mean structure of $y$ and $\epsilon$ is the zero-mean noise. As discussed in \citet{Buja2019}, the best population linear approximation to $y$ using only the features $\x$ can be written, assuming $\Ex(\x\x\Tr)$ is full rank, as 
\as{
\tilde{\bb} = \Ex(\x\x \Tr)^{-1}\Ex\left(\x\mu(\x, \z)\right). 
}

If the true mean structure is linear such that $\mu(\x, \z) = \x\Tr\bb + \z\Tr\bg$, then the only source of misspecification comes from omitting $\z$ from the linear model, thereby projecting $\mu(\x, \z)$ onto $\x$. In this case, $\tilde{\bb}$ can be written as 
\als{beta_tilde}{
\tilde{\bb} &= \Ex(\x\x\Tr)^{-1}\Ex\left(\x(\x\Tr\bb + \z\Tr\bg)\right) \\
&= \bb + \underbrace{\Ex(\x\x\Tr)^{-1}\Ex(\x\z\Tr)\bg} \\
&= \bb + \hspace{13.5mm}\btau. 
}
Equation \eqref{beta_tilde} shows that in the linear regression setting, unobserved confounders introduce bias $\btau=\Ex(\x\x\Tr)^{-1}\Ex(\x\z\Tr)\bg$: as more data becomes available, the least squares estimator $\bbh$ converges to $\bb + \btau$, not $\bb$. The bias $\btau$ can also be interpreted as the extent to which the confounding effects $\z\Tr\bg$ can be projected onto the nearest distribution in the working model.

Using the definition above and under the linearity assumption, the response random variable $y$ can be decomposed as 
\als{y_decomp}{
y &= \hspace{7mm} \x\Tr\tilde{\bb} \hspace{2.5mm} + \left( \hspace{4mm} \mu(\x, \z) \hspace{3.8mm} - \hspace{7mm} \x\Tr\tilde{\bb}\right) \hspace{7.5mm} + \epsilon \\
&= \x\Tr\overbrace{(\bb+\btau)} + \left(\overbrace{\x\Tr\bb + \z\Tr\bg}-\overbrace{(\x\Tr\bb+\x\Tr\btau)}\right) \hspace{1.5mm} + \epsilon \\
&= \x\Tr(\bb+\btau) \hspace{0.5mm} + \hspace{12mm} \underbrace{(\z\Tr\bg-\x\Tr\btau)} \hspace{16mm} + \hspace{1mm} \epsilon \\
&= \x\Tr(\bb+\btau) \hspace{0.5mm} + \hspace{21mm} \psi \hspace{24mm} + \hspace{0.2mm} \epsilon. 
}
Equation \eqref{y_decomp} illustrates the full impact of unobserved confounding on the response random variable. When the confounders $\z$ are omitted from the model, there are two consequences: in addition to the bias term $\btau$ discussed already, the part of $\z \Tr \bg$ that cannot be projected onto the distribution of $\x$ adds additional noise $\psi$ into the model, and this noise is not correlated with $\x$ (i.e., it is exogenous).

As in \citet{Buja2019}, equation \eqref{beta_tilde} shows that when a model is misspecified, its coefficient estimates become dependent on the distribution of both the features $\x$ and the confounders $\z$. This dependence is summarized by $\btau$, which depends on both $\Ex(\x\x\Tr)$ and $\Ex(\x\z\Tr)$. Note that if $\x$ and $\z$ are uncorrelated, then $\btau = \zero$ and all confounding effects $\z\Tr\bg$ contribute to noise $\psi$ without introducing bias. In the next section, we introduce a linear structure to define the relationship between $\x$ and $\z$. 

\subsection{Linear Confounding Model} \label{Sec:linear confounding model}

In the previous section, we demonstrated that the bias $\tau$ introduced by confounding depends on the joint distribution of $\x$ and $\z$. In this section, we propose a linear relationship between $\x$ and $\z$, which allows us to derive $\btau$ in closed form.

Let $\d$ denote a random vector of length $p$, with $\d$ independent of $\z$, and $\A$ be a fixed $p\times q$ matrix. Consider the following data generating mechanism, which we refer to as the \emph{linear confounding model}:
\als{Eq:setup}{
y &= \x\Tr\bb + \z\Tr\bg + \epsilon, \\
\x &= \d+\A\z, \\
\d, \z &\sim \Norm(\zero, \I_p), \quad \epsilon \sim \Norm(0, \sigma_e^2),
}
where $\d$, $\z$, and $\epsilon$ are mutually independent. In this model $\A$ determines the correlation between the confounders $\z$ and features $\x$. Note that the confounders influence both the features and the response. Since $\z$ are drawn from a standard normal distribution, we have $\Ex(\z\z\Tr)=\Var(\z)=\I_p$. Combining equations \eqref{beta_tilde} and \eqref{Eq:setup},
\als{Eq:tau_model}{
  \btau &=\Ex(\x\x\Tr)^{-1}\Ex(\x\z\Tr)\bg \\
  &= (\I_p + \A\Ex(\z\z\Tr)\A\Tr)^{-1}\A\Ex(\z\z\Tr)\bg \\
  &= (\I_p + \A\A\Tr)^{-1}\A\bg. 
}
Likewise, the additional noise added to the model by the confounders is
\als{Eq:var_psi}{
  \Var(\psi \given \btau)
  &= \bg\Tr\Var(\z)\bg + \btau\Tr\Var(\x)\btau - 2\bg\Tr\Var(\z)\A\Tr\btau \\
  &= \bg \Tr \bg + \btau \Tr (\I_p + \A\A \Tr )\btau - 2\bg \Tr \A \Tr \btau
  . 
}

The linear confounding assumption that $\z$ follows a standard Gaussian distribution, and thus that the confounders are independent, is not as restrictive as it seems. Suppose that we are in the presence of batch effects and $\z$ indicates which batch an instance came from; for example, $\z\Tr = [0 \quad 1 \quad 0]$ for an instance in the second of three batches. Clearly, $\z$ is no longer Gaussian; furthermore, $\Ex(\z\z\Tr) \ne \I_p$. However, there exists an equivalent linear confounding model with different values of $\A$ and $\bg$. Specifically, if we rotate $\A$ and $\bg$ by $\left\{\Ex(\z\z\Tr)\right\}^{1/2}$, we can compute $\btau$ from equation \eqref{Eq:tau_model} as 
$$ \btau = (\I_p + \Tilde{\A}\Tilde{\A}\Tr)^{-1}\Tilde{\A}\Tilde{\bg}, $$
where $\Tilde{\A} = \A\left\{\Ex(\z\z\Tr)\right\}^{1/2}$ and $\Tilde{\bg} = \left\{\Ex(\z\z\Tr)\right\}^{1/2}\bg$. In other words, since the covariance structure of $\z$ can be absorbed into $\A$ and $\bg$, it suffices to consider the $\z \sim \Norm(\zero, \I)$ case without loss of generality. The linear confounding model is easily modified to produce features that are standardized (mean zero, unit variance); see supplemental material.

\subsection{Decomposition of Signal, Bias, and Noise} \label{Sec:quants}

In simulation studies, the signal-to-noise ratio is a standard quantity that characterizes the difficulty of the problem. In our setting, the presence of unobserved confounding introduces the additional complication of bias. To account for this more complex situation, below we define two additional measures: the bias-to-noise ratio and the bias-to-signal ratio. Explicitly defining these quantities allows us to vary one while holding others fixed --- for example, adjusting the bias-to-signal ratio while keeping the signal-to-noise ratio constant. We have found that the ability to isolate effects in this manner is critical to designing clear, interpretable simulations.

Section~\ref{Sec:misspecification} illustrated how unobserved confounding can be decomposed into bias and noise. Given the quantities introduced in equations \eqref{y_decomp}, \eqref{Eq:tau_model} and \eqref{Eq:var_psi}, we define: 
\als{Eq:quantities}{
    \text{Signal-to-Noise Ratio } &= \frac{\bb\Tr\Var(\x)\bb}{\Var(\psi\given\btau) + 1}, \\
    \text{Bias-to-Signal Ratio } &= \frac{\btau\Tr\Var(\x)\btau}{\bb\Tr\Var(\x)\bb}, \\
    \text{Bias-to-Noise Ratio } &= \frac{\btau\Tr\Var(\x)\btau}{\Var(\psi\given\btau) + 1}. \\
}
The Signal-to-Noise Ratio (SNR) quantifies the magnitude of the signal relative to the total noise (from both error and confounding) in $\y$. The Bias-to-Signal Ratio (BSR) measures the magnitude of the bias relative to the magnitude of the signal. Lastly, the Bias-to-Noise Ratio (BNR) quantifies the relative magnitude of the bias compared to the noise. With unobserved confounding, both SNR and BNR contribute to the difficulty of the problem. A larger SNR will tend to make the problem easier, but a problem with a high SNR may be difficult if BNR is also high. Note that specifying any two of these three quantities automatically determines the third one. Together, they facilitate the construction of simulation studies and the interpretation of simulation results. 


\section{Simulation} \label{Sec:sim}

In this section, we present simulation studies comparing the performance of LASSO, PC-LASSO, and PLMM under different confounding scenarios. Values of the tuning parameter $\lambda$ are chosen based on the minimum cross-validation errors (CVE) for all three methods. 

\subsection{Simulation Design} \label{Sec:sim_parameters}

Throughout, all simulation studies are based on the linear confounding model \eqref{Eq:setup}. However, because these simulations involve variable selection and penalized regression, some care is needed in the construction of $\A$ to avoid misleading results. In particular, if the confounding were such that coefficients were always biased away from zero, this would, this bias would counteract the LASSO's tendency to shrink estimates toward zero. The two biases would cancel, producing a simulation that is skewed in favor of LASSO.

\begin{figure}[h]
    \centering
    \begin{tikzpicture}[scale=1.2]
        \draw[thick] (0,0) -- (4.5,0);
        \draw[thick] (0,-0.1) -- (0,0.1) node[below=6pt] {0};
        \draw[thick] (1.5,-0.1) -- (1.5,0.1) node[below=6pt] {$\beta-\delta$};
        \draw[thick] (4,-0.1) -- (4,0.1) node[below=6pt] {$\beta$};
        \draw[thick] (2.8,-0.1) -- (2.8,0.1) node[above=-2pt] {$\beta-\delta+\tau$};
        \draw [<-, thick] (1.5,-0.6) -- (2.7,-0.6); 
        \node at (2.8,-0.6) {\footnotesize $\delta$};
        \draw[thick] (2.9,-0.6) -- (4,-0.6);

        \draw [thick] (1.5,-1.0) -- (2.1,-1.0);
        \node at (2.2,-1.0) {\footnotesize $\tau$};
        \draw [->, thick] (2.3,-1.0) -- (2.8,-1.0);

        \draw [<-, ultra thick] (2.8,-1.4) -- (3.1,-1.4);
        \node at (3.43,-1.4) {\footnotesize $\delta-\tau$};
        \draw [ultra thick] (3.75,-1.4) -- (4,-1.4);
    \end{tikzpicture}
    \hspace{2cm}
    \begin{tikzpicture}[scale=1.2]
        \draw[thick] (0,0) -- (4.5,0);
        \draw[thick] (0,-0.1) -- (0,0.1) node[below=6pt] {0};
        \draw[thick] (1,-0.1) -- (1,0.1) node[below=6pt] {$\beta-\delta-\tau$};
        \draw[thick] (4,-0.1) -- (4,0.1) node[below=6pt] {$\beta$};
        \draw[thick] (2,-0.1) -- (2,0.1) node[above=-2pt] {$\beta-\delta$};
        \draw[<-, thick] (2,-0.6) -- (2.9,-0.6);
        \node at (3, -0.6) {\footnotesize $\delta$};
        \draw[thick] (3.1,-0.6) -- (4,-0.6);
        
        \draw[<-, thick] (1,-1.0) -- (1.4,-1.0);
        \node at (1.5, -1.0) {\footnotesize $\tau$};
        \draw[thick] (1.6,-1.0) -- (2,-1.0);
        
        \draw[<-, ultra thick] (1,-1.4) -- (2.2,-1.4);
        \node at (2.5, -1.4) {\footnotesize $\delta+\tau$};
        \draw[ultra thick] (2.8,-1.4) -- (4,-1.4);
    \end{tikzpicture}
    \caption{The left panel shows that if $\btau$ is away from zero, the overall bias magnitude is $\btau-\bd$. The right panel shows that if $\btau$ is towards zero, the overall bias magnitude is $\btau+\bd$. }
    \label{Fig:same sign tau}
\end{figure}
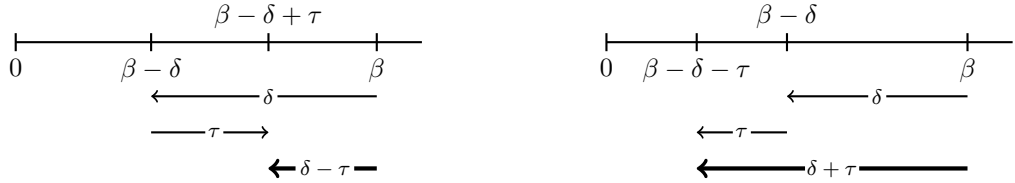

To resolve this issue, both positive and negative biases need to be incorporated for a more balanced and realistic assessment of LASSO’s estimation performance. Consider the case where the true signals in $\bb$ are all positive, and let $\bd$ denote LASSO's shrinkage toward zero. Figure \ref{Fig:same sign tau} depicts two scenarios for a single signal $\beta$, depending on the direction of the bias $\tau$ introduced by confounding. In the left panel, the confounding bias $\tau$ is positive, pushing the estimate away from zero. The combined effect of confounding and LASSO yields an estimate of $\beta-\delta+\tau$, with total bias $\delta-\tau$. Here, the confounding bias partially offsets LASSO's shrinkage, thereby making LASSO look better. In the right panel, $\tau$ is in the same direction as LASSO's shrinkage, pushing the estimate closer to zero. The resulting estimate becomes $\beta-\delta-\tau$, and the total bias is $\delta+\tau$. In this case, the two biases compound, leading to a worse estimate for LASSO. Neither scenario alone accurately captures LASSO's performance in the presence of confounding. To ensure a fair comparison with other methods, the simulation must balance positive and negative bias across features so that the LASSO's shrinkage does not create systematic advantages or disadvantages.

The following construction of $\A$ avoids the above phenomenon. To illustrate, suppose there are four features, two confounders, and 
\as{
\A = \begin{bmatrix}
    a & 0 \\
    -a & 0 \\
    0 & a \\
    0 & -a \\
\end{bmatrix}.}
With this setup, the correlation structure of $\x$ is
\as{
\begin{bmatrix}
    1 & -\rho & 0 & 0 \\
    -\rho & 1 & 0 & 0 \\
    0 & 0 & 1 & -\rho \\
    0 & 0 & -\rho & 1 \\
\end{bmatrix}, 
}
where $\rho=a^2/(1+a^2)$. 

In this configuration, the first two features are influenced by the first confounder, while the remaining two are influenced by the second. Features affected by the same confounder are correlated, with parameter $a$ controlling the strength of these correlations. The presence of both positive and negative entries in $\A$ induces a mixture of positive and negative bias in $\btau$. For  the $4\times 2$ matrix $\A$ in this example and $\bg\Tr=\begin{bmatrix}
    g & g
\end{bmatrix}$, we can apply equation \eqref{Eq:tau_model} to see that $\btau$ is evenly balanced between positive and negative biases:
\as{ \btau\Tr = 
\frac{ag}{1 + 2a^2}
\begin{bmatrix}
    1 & -1 & 1 & -1 \\
\end{bmatrix}.
}

This pattern can be generalized to incorporate more features and confounders. For example, if there are 8 features and 2 confounders, then $\A$ is the $8\times 2$ matrix 
\as{\A\Tr=
    \begin{bmatrix}
        a & a & -a & -a & 0 & 0 & 0 & 0 \\
        0 & 0 & 0 & 0 & a & a & -a & -a \\
    \end{bmatrix},
}
indicating that the first 4 features are affected by the first confounder. The first 2 of these 4 features are positively correlated with the confounder, while the other 2 features are negatively correlated with the confounder. The same idea applies to the other 4 features, which are affected by the second confounder. The correlation structure of $\x$ is therefore an $8\times 8$ block-diagonal matrix, where the two off-diagonal matrices are all 0 and the two $4\times 4$ matrices on the diagonal are
\as{
    \begin{bmatrix}
        1 & \rho & -\rho & -\rho \\
        \rho & 1 & -\rho & -\rho \\
        -\rho & -\rho & 1 & \rho \\
        -\rho & -\rho & \rho & 1 \\
    \end{bmatrix}. 
}
As before, $\btau$ is an equal mixture of positive and negative biases, although here, both positive and negative correlations exist among the features. For a general $p\times q$ block-diagonal matrix $\A$, each column of $\A$ has $m = p / q$ nonzero entries, half of which equal $a$, the other half equal to $-a$. The remaining $p-m$ entries are zero, indicating that those features are unaffected by the confounder.

With this design of $\A$, if all elements of $\bg$ are equal to $g$, then $\btau$ and $\Var(\psi\given\btau)$ are functions of $a$ and $g$. However, the magnitude of the bias is not a monotone function of $a$ --- multiple values of $a$ and $g$ correspond to the same $\btau\Tr\btau$. To fix this, we reparameterize by specifying the ratio $g = ra$ (this causes $\norm{\btau}$ to be a monotone function of $a$ and $r$; see Appendix for details).

This parameterization above allows us to fix $\Var(\psi\given\btau)=1$ throughout and specify the value of BNR, then solve for $a$ and $r$. Given these, and assuming $\bb$ is a vector of length $p$ with a specified number of nonzero elements equal to $b$, the signal effect size $b$ is then determined by the SNR. 

In simulation studies, we kept $\Var(\psi\given\btau)=1, n=300, p=600$ and specified 8 true signals that were randomly chosen among all 600 features. We utilized SNR, BNR, and BSR to investigate how the amount and complexity of unobserved confounding affects model performance. PC-LASSO models were fit using 10 PCs, unless specified otherwise. 

\subsection{Magnitude of confounding}

In this section, we examine how varying amounts of confounding affect the performance of different models, while the confounding structure remains the same. As introduced in Section \ref{Sec:linear confounding model}, the confounding effect can be decomposed into two parts --- bias and noise. Since an increase in noise generally degrades the performance of all models, our focus here is specifically on the bias introduced by confounding. In this section therefore, we fix SNR = 1.5 and vary the BNR from 0 to 3. All simulations involve $q=10$ unmeasured confounders and results are averaged over 1,000 replications; see Table~\ref{Tab:changing amount table} for details.

\begin{table}[ht]
\centering
\begin{tabular}{rrrrrrr}
  \hline
  q & BNR & BSR & SNR & $\rho$ & r & b \\ 
  \hline
  10 & 0.00 & 0.00 & 1.50 & 0.00 & 0.00 & 0.43 \\ 
  10 & 0.50 & 0.33 & 1.50 & 0.02 & 3.46 & 0.61 \\ 
  10 & 1.50 & 1.00 & 1.50 & 0.05 & 2.83 & 0.62 \\ 
  10 & 3.00 & 2.00 & 1.50 & 0.09 & 2.65 & 0.63 \\ 
  \hline
\end{tabular}
\caption{Parameter values used in simulations to vary the amount of bias by changing BNR while keeping SNR=1.5 and $\Var(\psi\given \btau)=1$. When BNR=0, $\Var(\psi\given \btau)=0$, corresponding to no confounding at all. Ten confounders were used ($q=10$). }
\label{Tab:changing amount table}
\end{table}

\begin{figure}[ht]
    \centering
    \includegraphics[scale=1]{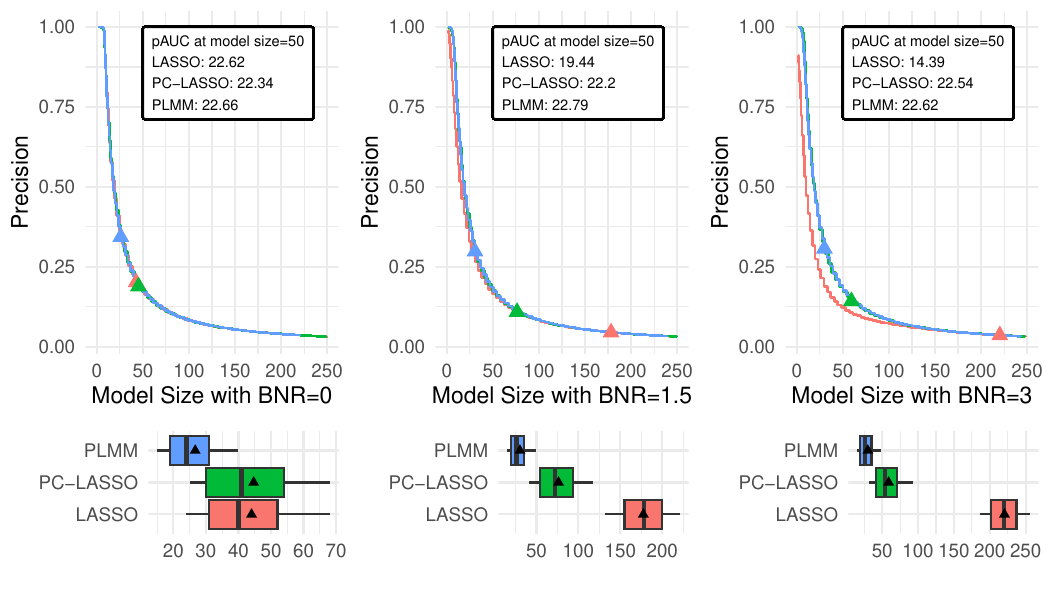}
    \caption{Precision curves of the three models with different amounts of bias, reflected by three levels of BNR. The top panels display the precision curves as the model sizes change. Triangles on each curve indicate the precision at the average model size selected via CV for each method. The bottom panels show boxplots of the model sizes for each method. Black triangles denote the averaged model sizes. To reduce the impact of extreme values, the whiskers of the boxplots extend to the 0.1 and 0.9 quantiles instead of the minimum and maximum.}
    \label{Fig:bnr_precision}
\end{figure}

Figure \ref{Fig:bnr_precision} illustrates how feature selection precision and model size change with varying BNR. In the feature selection precision curve, the horizontal axis represents the number of non-zero coefficients in each model (i.e., model size), while the vertical axis is
\as{
\text{Precision} = \frac{\text{True Positive}}{\text{True Positive} + \text{False Positive}}. 
}
As model size increases, more noise features get selected, reducing precision due to a rise in false positives. For each curve, the partial area under the curve (pAUC) is computed as the area under the precision curve up to a model size of 50. A higher pAUC indicates better performance in identifying true signals with smaller models.

The top panel shows that when BNR is 0 (no bias due to confounding), all three models achieve comparable precision, reflected in similar pAUC values. However, as BNR increases, the negative impact of unadjusted confounding becomes more pronounced. The pAUC of LASSO declines from 22.6 to 14.4 as BNR increases from 0 to 3. By failing to account for unobserved confounding, the LASSO selects a large number of false positives --- features that are not causally related to the outcome, but merely associated due to confounding. Throughout, LASSO demonstrated a consistent tendency to select a large number of false positives.

In contrast, PC-LASSO and PLMM maintain relatively stable precision across all BNR levels. This robustness is somewhat surprising --- the methods select features just as accurately when confounding is present as when it is absent. However, despite their similar selection precision, the two adjusted methods differ in model size. The bottom panel of Figure \ref{Fig:bnr_precision} shows that PLMM consistently selects more compact models than PC-LASSO, achieving strong precision with fewer features. 

\begin{figure}[ht]
    \centering
    \includegraphics[scale=0.8]{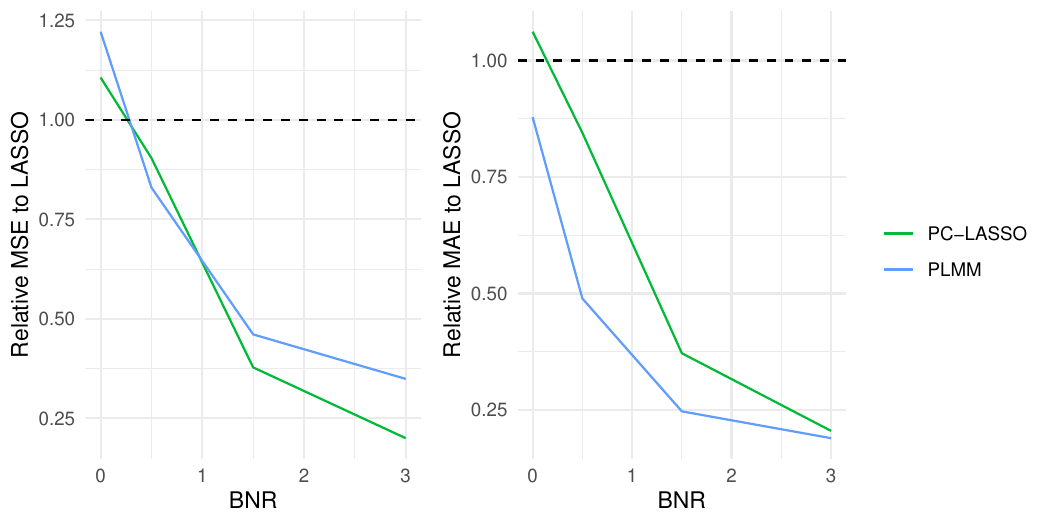}
    \caption{Relative mean squared error (MSE) and mean absolute error (MAE) compared to LASSO as BNR changes. \emph{Left:} MSE $=\Ex(\norm{\hat{\bb}-\bb}_2^2)$ relative to LASSO. \emph{Right:} MAE $=\Ex(\norm{\hat{\bb}-\bb}_1)$ relative to LASSO. The green line represents PC-LASSO, and the blue line represents PLMM. The dashed line is reference at 1. }
    \label{Fig:bnr_msemae_line}
\end{figure}

Figure \ref{Fig:bnr_msemae_line} compares estimation performance relative to LASSO. When BNR is small, both PC-LASSO and PLMM have relative MSE and MAE above 1, implying LASSO is outperforming these methods slightly --- this not surprising considering that there is little need to adjust for confounding in these scenarios. However, as BNR increases, both PC-LASSO and PLMM show increasing advantages over LASSO, with MSE and MAE roughly 4 times smaller when BNR = 3.

Overall, PC-LASSO yields the lowest MSE, whereas PLMM performs best with respect to MAE. We found this pattern to hold consistently throughout all of the simulation scenarios. These differences reflect the trade-offs inherent in the evaluation metrics: MSE penalizes large errors more heavily, whereas MAE is more robust to error magnitudes and rewards PLMM's more parsimonious models (fewer but larger errors).

\subsection{Number of Features} 

In the previous section, we saw that PC-LASSO and PLMM were essentially unaffected by the presence of unmeasured confounders. This would seem to contradict classical statistical theory, which indicates that unobserved confounders should have a negative impact on estimation and variable selection. Here, we show that these methods are only able to adjust for unmeasured confounding in high-dimensional scenarios.

In Section \ref{Sec:methods}, we discussed how both PC-LASSO and PLMM adjust for unobserved confounding using information from the observed covariate matrix $\X$. PC-LASSO achieves this by including the first $k$ PCs of $\X$ as unpenalized covariates, while PLMM models the confounders as random effects and estimates their variance-covariance structure based on $\X$. In this section, we present a simple visualization to illustrate that, in order to adequately account for varying levels of unobserved confounding, both methods require a large number of features, which enables a more accurate approximation of the unobserved confounding. 

\begin{figure}[ht]
    \centering
    \includegraphics[scale=0.8]{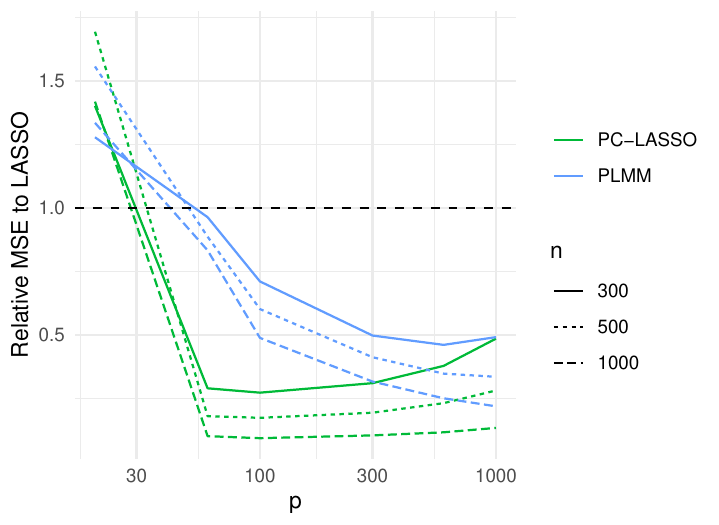}
    \caption{Relative MSE of PC-LASSO and PLMM to LASSO changing with $p$ using various sample sizes $n$. }
    \label{Fig:pMSE}
\end{figure}

Figure \ref{Fig:pMSE} is generated by fixing these settings while allowing $p$ and $n$ to vary: BNR=1.5, BSR=1, SNR=1.5, and $\Var(\psi\given \btau)=1$ with $q=10$ confounders. PC-LASSO models are fitted using 10 PCs. The figure illustrates that when the number of features $p$ is small, both PC-LASSO and PLMM exhibit higher MSE than LASSO, regardless of the sample size. As $p$ increases, both methods are better at approximating the unobserved confounding structure, leading to a decline in relative MSE, which eventually drops below 1. While larger sample sizes contribute to improved approximation and better estimation, they cannot compensate for insufficient features. That is, if $p$ is too small, a large amount of data cannot fully recover the confounding structure, limiting the effectiveness of adjustment. This observation is consistent with the conclusion drawn by \citet{Hayes2009} that a large number of markers is required for a reliable prediction, and also with the findings of \citet{Wang2019}.

In Figure~\ref{Fig:pMSE}, PC-LASSO demonstrates better estimation accuracy. This may be attributed to two factors. First, the number of unobserved confounders is relatively small ($q=10$), making it easier for a few PCs to capture the underlying confounding structure and separate it from the signals. Second, the number of PCs used in the model matches the number of confounders, giving PC-LASSO an added advantage by providing sufficient adjustment without overfitting. In the following section, we further explore how the number of unobserved confounders impacts model performance and examine the selection of PCs in PC-LASSO models.

\subsection{Complexity of Confounding} \label{Sec:complexity}

In this section, we show that in addition to the \emph{amount} of confounding, its \emph{complexity}, characterized by the number of unobserved confounders $q$, plays a crucial role in model performance. A larger $q$ corresponds to a more complex confounding structure, as more latent factors influence the data. To isolate the effect of complexity, we fixed BNR at 1.5, BSR at 1, SNR at 1.5, and set $\Var(\psi \given \btau) = 1$ (see Table \ref{Tab:changing q table}).

\begin{table}[ht]
\centering
\begin{tabular}{rrrrrrr}
  \hline
  q & BNR & BSR & SNR & $\rho$ & r & b \\ 
  \hline
    2 & 1.50 & 1.00 & 1.50 & 0.01 & 14.14 & 0.61 \\ 
    3 & 1.50 & 1.00 & 1.50 & 0.01 & 9.43 & 0.61 \\ 
    5 & 1.50 & 1.00 & 1.50 & 0.02 & 5.66 & 0.61 \\ 
    6 & 1.50 & 1.00 & 1.50 & 0.03 & 4.71 & 0.62 \\ 
    10 & 1.50 & 1.00 & 1.50 & 0.05 & 2.83 & 0.62 \\ 
    20 & 1.50 & 1.00 & 1.50 & 0.09 & 1.41 & 0.61 \\ 
    60 & 1.50 & 1.00 & 1.50 & 0.23 & 0.47 & 0.63 \\ 
    100 & 1.50 & 1.00 & 1.50 & 0.33 & 0.28 & 0.64 \\ 
    150 & 1.50 & 1.00 & 1.50 & 0.43 & 0.19 & 0.61 \\ 
   \hline
\end{tabular}
\caption{Parameter values used in simulations to vary the complexity of confounding, keeping BNR=1.5, BSR=1, SNR=1.5, and $\Var(\psi\given \btau)=1$. } 
\label{Tab:changing q table}
\end{table}

\begin{figure}[ht]
    \centering
    \includegraphics[scale=0.8]{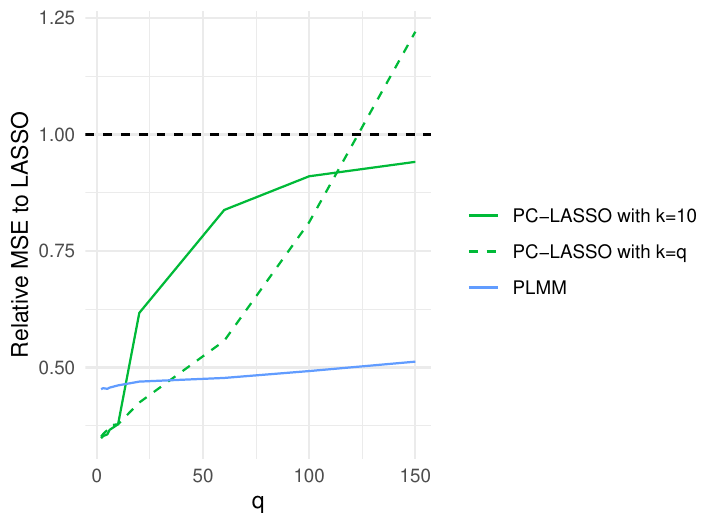}
    \caption{Relative MSE compared to LASSO as $q$ changes. The two green lines represent PC-LASSO with different choices of $k$, and the blue line represents PLMM.}
    \label{Fig:q_mse}
\end{figure}

Figure \ref{Fig:q_mse} illustrates the impact of confounding complexity on estimation accuracy. PC-LASSO models were fit using both $k=10$ and $k=q$ (the true number of confounders). Initially, when $q$ is small, PC-LASSO achieves lower MSE than PLMM, regardless of the choice of $k$. However, as $q$ increases, PC-LASSO's MSE increases rapidly. When using $k=10$, the relative MSE of PC-LASSO grows fast at first but eventually stabilizes and is never worse than LASSO. When using $k=q$, PC-LASSO is able to maintain lower MSE than PLMM for small to moderate $q$, but then the MSE increases sharply and eventually exceeds the MSE of LASSO. In contrast, PLMM demonstrates stable MSE across all values of $q$, with an MSE roughly 50\% lower than both PC-LASSO and LASSO when confounding is most complex.

In summary, PC-LASSO's performance is highly related to the number of PCs in the model. It performs well under simple confounding structures with fewer confounders but struggles as the number of unobserved confounders grows. PLMM, on the other hand, maintains stable performance.

Having established PC-LASSO’s sensitivity to the choice of $k$, we now take a closer look at how varying $k$ influences model performance across different confounding scenarios. We begin with a baseline scenario where $q=10$, representing a moderate confounding complexity. To assess the impact of $k$ under more complex confounding, we consider a second scenario with $q=150$. Finally, in the third scenario, we study the case where only a subset of the unobserved factors influences the outcome. Specifically, 20 unobserved factors affect $\X$, but only half of them also affect $\y$, making those 10 ``true'' confounders. In this case, $\bg$ is of length 20, with 10 nonzero entries. Without loss of generality, we set the first ten unobserved factors as confounders and define $\bg\Tr=(ra, ..., ra, 0, ..., 0)$. All three scenarios were constructed fixing $\Var(\psi\given\btau)=1$, BNR=1.5, SNR=1.5, and BSR=1 to ensure comparability across settings.

\begin{figure}[ht]
    \centering
    \includegraphics[scale=0.85]{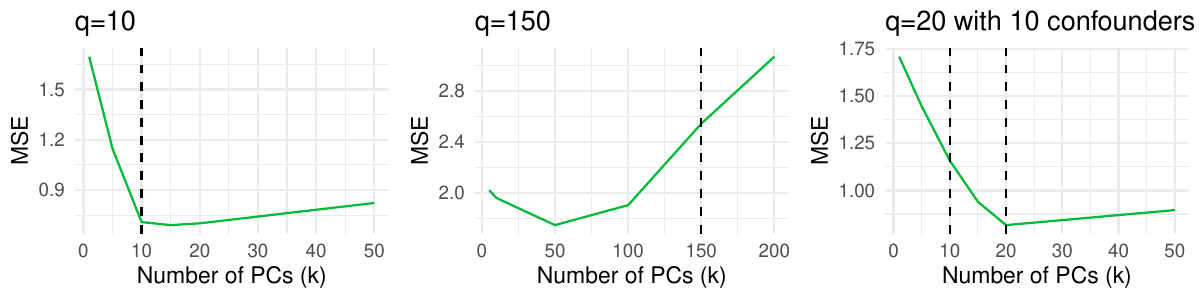}
    \caption{Effect of the choice of $k$ in PC-LASSO models under three different scenarios. Dashed lines indicate the number of confounders (and unobserved factors) in each scenario. }
    \label{Fig:pcnum_mse}
\end{figure}

Figure \ref{Fig:pcnum_mse} illustrates how the MSE varies with different choices of $k$ across the three scenarios described earlier. In the left panel, where $q=10$, the lowest MSE is achieved when $k=10$, matching the number of confounders. Moreover, the model is more sensitive to under-adjustment (too few PCs) than to over-adjustment (too many PCs). In the middle panel, where $q=150$, the smallest MSE is no longer achieved when $k=q$. Here, adding large numbers of parameters to the model greatly increases variability and does more harm than good in terms of adjustment. Finally, in the right panel, which considers the hybrid case, the smallest MSE is achieved when $k=20$, implying that adjusting for all latent structure in $\X$, even non-confounding factors, can still aid estimation.

These findings underscore that while $k=10$ is a reasonable default, there is no universally optimal number of PCs to include, as the best value depends on the confounding structure. In practice, since the true number and structure of confounders are unknown, selecting $k$ is nontrivial.


\section{Study of gene expression in breast cancer patients}

In this section, we examine the breast cancer subset of the TCGA Research Network (\url{https://www.cancer.gov/tcga}), which includes clinical data and expression profiles of 19,962 genes for $n=1,231$ patients who were diagnosed with breast cancer. All expression measures were log transformed prior to analysis. We treat BRCA1 gene expression as the outcome --- reflecting its well-established role in breast cancer --- and the remaining $p=19,961$ genes as predictors. Our goal is to identify genes whose expression patterns are associated with BRCA1, and which may provide further insights into breast cancer biology.

\subsection{Semi-synthetic simulation}

In Section \ref{Sec:sim}, we examined various scenarios under the linear confounding model in order to isolate the effects of unobserved confounding. With real data, confounding is likely to be more ``messy.'' To explore a more realistic scenario, we carried out a simulation using features from the TCGA breast cancer data. As a confounder, we used previously defined breast cancer subtypes based on protein-expression profiling \citep{Koboldt2012}. Each sample belongs to one of five such subtypes (plus a sixth subtype for samples in which proteomic data was missing), which are strongly associated with broad patterns in the gene expression data. In our simulation, this subtype variable was specified to have a direct effect on the simulated outcome while also being correlated with the observed gene expression features. The expression data itself remained unchanged; only the outcome was simulated. This setup provided a semi-synthetic benchmark in which the gene expression matrix retained real biological structure while the outcome was influenced by both causal genes and a realistic confounder.

Letting $\Z$ denote a 1231$\times$6 matrix indicating subtype, we generated $\y = \X\bb+\Z\bg+\bvep$, where $\bg$ is a vector of length 6, and $\bvep$ is random Gaussian noise. Five randomly selected entries of $\bb$ were set to 1 with the others equal to 0, while $\bg = (-g, -g, -g, g, g, g)$ and we vary the impact of the confounder by changing the magnitude of $g$. Results are averaged over 100 replications.

\begin{table}[ht]
\centering
\begin{tabular}{rrrrr}
  \hline
    $g$ & 0 & 0.5 & 1 & 2 \\
  \hline
    \multicolumn{4}{l}{\textbf{Relative MSE to LASSO}} \\
    \quad PC-LASSO & 2.07 & 1.70 & 1.42 & 1.19 \\ 
    \quad PLMM     & 1.29 & 1.07 & 0.86 & 0.79 \\ 
    \multicolumn{4}{l}{\textbf{Relative MAE to LASSO}} \\
    \quad PC-LASSO & 1.50 & 1.25 & 1.05 & 0.92 \\ 
    \quad PLMM     & 0.86 & 0.68 & 0.53 & 0.48 \\ 
    \multicolumn{4}{l}{\textbf{pAUC at Model Size 50}} \\
    \quad LASSO    & 12.57 & 11.91 & 10.25 & 7.06 \\ 
    \quad PC-LASSO & 34.33 & 33.76 & 31.72 & 24.26 \\ 
    \quad PLMM     & 37.82 & 37.86 & 37.31 & 33.71 \\ 
    \multicolumn{4}{l}{\textbf{Model Size}} \\
    \quad LASSO    & 56.74 & 78.41 & 104.65 & 118.57 \\ 
    \quad PC-LASSO & 74.39 & 79.26 & 85.88 & 85.59 \\ 
    \quad PLMM     & 26.28 & 24.89 & 23.89 & 25.35 \\ 
   \hline
\end{tabular}
\caption{Relative MSE and MAE to those of LASSO when $\X$ and $\Z$ are from the real data. Relative MSE or MAE values $>$ 1 indicate worse performance than LASSO. Averaged pAUC and model sizes are used to summarize the performance in feature selection.}
\label{Tab:semi-real performance res}
\end{table}

Table \ref{Tab:semi-real performance res} summarizes the performance of the three methods with varying $g: \{0, 0.5, 1, 2\}$. As $g$ increases, both PC-LASSO and PLMM improve (relative to the LASSO) in both MSE and MAE, emphasizing the importance of adjustment when confounding is present. Moreover, PLMM noticeably outperforms LASSO in terms of estimation accuracy when there is a large amount of confounding. PC-LASSO, on the other hand, does not, suggesting that principal components may fail to represent the complexity of real $\X$ matrices ($k=10$).

In terms of variable selection, however, LASSO consistently exhibits the worst performance, perhaps because it struggles to select the correct variable in the presence of correlated features \citep{Zou2005}. While PC-LASSO performs much better than LASSO, PLMM has the highest accuracy in terms of variable selection across all levels of $g$, while also producing the most parsimonious models. While the pAUC of all three models declines with increasing $g$, PLMM is the most robust to increasing amounts of confounding.

\subsection{Real Data Analysis}

For this analysis, we fit LASSO, PC-LASSO with different choices of $k$, and PLMM models with BRCA1 expression as the outcome. For each method, a 10-fold CV was used to select the tuning parameter $\lambda$. All models used $\lambda$ values that are associated with the lowest CVE. Table \ref{Tab:real data analysis} summarizes the prediction error (PE) and model size for each method, along with the null model as a reference.

Unsurprisingly, all models greatly outperform the null model. The LASSO has a higher PE than the PC-LASSO and PLMM models and selects a large model with 249 genes. PC-LASSO's performance depends on the number of PCs included in the model. When $k$ is small, there is relatively less adjustment for the unobserved confounding effect and models similar in size to the LASSO, with over 200 genes included. PLMM, on the other hand, selects a much smaller model with only 36 genes, while matching PC-LASSO for the best prediction accuracy.

\begin{table}[ht]
\centering
\begin{tabular}{lrrr}
    \hline
    Method & Model Size & PE & Cancer Genes \% \\ 
    \hline
    Null Model & 0 & 0.67 & \\ 
    LASSO & 249 & 0.21 & 59 \\ 
    PC-LASSO with $k=3$ & 260 & 0.18 & 58 \\ 
    PC-LASSO with $k=10$ & 261 & 0.18 & 57 \\ 
    PC-LASSO with $k=20$ & 265 & 0.19 & 58 \\ 
    PC-LASSO with $k=50$ & 125 & 0.19 & 58 \\ 
    PLMM & 36 & 0.18 & 81 \\   
    \hline
\end{tabular}
\caption{Prediction error (PR) and model size for all models selected by 10-fold CV} 
\label{Tab:real data analysis}
\end{table}

As this is real data, we cannot know which genes have a true biological association with BRCA1, but as a proxy, we assembled a broad list of cancer-related genes from the MSigDB C2 collection whose names included either ``cancer'' or ``tumor.'' Since BRCA1 is a tumor suppressor, genes that modify the expression of BRCA1 would be more likely to show up on this list. According to this criterion, 81\% the genes selected by PLMM are plausibly related to cancer, a much higher percentage than any other method.


\section{Conclusion}

In this paper, we proposed a linear confounding model that allows decomposition and explicit control over the amount and complexity of unobserved confounding effects. This enables systematic evaluation of model performance under various confounding scenarios, which provided novel insights into the scenarios in which various approach to deconfounding work best. While we used the linear confounding model here to compare PC-LASSO and PLMM, the framework also lays a foundation for future studies of other methods of adjusting for unobserved confounding.

We carried out extensive simulations using this model and showed that with high-dimensional data, PC-LASSO and PLMM are both effective ways to adjust for unobserved confounding. Both approaches selected fewer false positives and resulted in more accurate estimates than a standard LASSO analysis in situations where confounding is present. Furthermore, we identified scenarios in which method outperforms the other. Overall, PLMM is more robust to the complexity of the confounding structure, as it does not rely on a specific low-dimensional representation. This was particularly noticeable in the analyses involving real data. In addition, while PC-LASSO and PLMM both prevent the inclusion of spurious features in the regression model, PLMM consistently produces more parsimonious models.

\bibliographystyle{ims-nourl}

\newpage

\begin{appendices}

\section{Standardization in Linear Confounding Model} 
For the linear confounding model described in \eqref{Eq:setup}, we will have $\Ex(\x) = \zero$, but the variance of $\x$ will not be $\one$. Here, we describe a modified version of the linear confounding model that produces a standardized matrix of features.

The variance-covariance matrix of $\x$ is given by 
$$ \Var(\x) = \Var(\d+\A\z) = \I_p + \A\Var(\z)\A\Tr. $$
Let $\S$ be a $p\times p$ diagonal matrix with elements $s_{ii} = (v_{ii})^{-1/2}$, where $v_{ii}$ is the corresponding element of $\Var(\x)$; i.e., the variance of feature $i$. Pre-multiplying $\x$ by $\S$ so the variance-covariance matrix of $\S\x$ is 
$$ \Var(\S\x) = \S\left(\I_p + \A\Var(\z)\A\Tr\right)\S, $$
where the diagonal elements are all 1. In other words, $\S\X$ is the standardized version of $\X$ (mean zero and unit variance). 

With the standardized features, we have 
\as{
\btau &= \left[\S(\I_p + \A\Ex(\z\z\Tr)\A\Tr)\S\Tr\right]^{-1}\S\A\Ex(\z\z\Tr)\bg \\
  &= \left[\S(\I_p + \A\A\Tr)\S\Tr\right]^{-1}\S\A\bg
}
if $\z$ follows a standard normal distribution and 
\as{
\btau &= \left[\S(\I_p + \Tilde{\A}\Tilde{\A}\Tr)\S\Tr\right]^{-1}\S\Tilde{\A}\Tilde{\bg}, 
}
if $\Ex(\z\z\Tr) \ne \I_p$, where $\Tilde{\A} = \A\left\{\Ex(\z\z\Tr)\right\}^{1/2}$ and $\Tilde{\bg} = \left\{\Ex(\z\z\Tr)\right\}^{1/2}\bg$. 


\section{Reparameterization for Simulation} 
We compute $\btau\Tr\btau$ to illustrate how $a$ and $g$ affect $\btau$. Based on equation \eqref{Eq:tau_model}, $\btau\Tr\btau$ can be computed as 
\as{
    \btau\Tr\btau = \frac{pa^2g^2(a^2+1)}{(ma^2+1)^2} = \frac{a^4(pg^2) + a^2(pg^2)}{a^4m^2 + a^2(2m) + 1} = \frac{a^4(pg^2) + a^2(pg^2)}{a^4(p/q)^2 + a^2(2p/q) + 1}. 
} 

\begin{figure}[ht]
    \centering
    \includegraphics[scale=0.75]{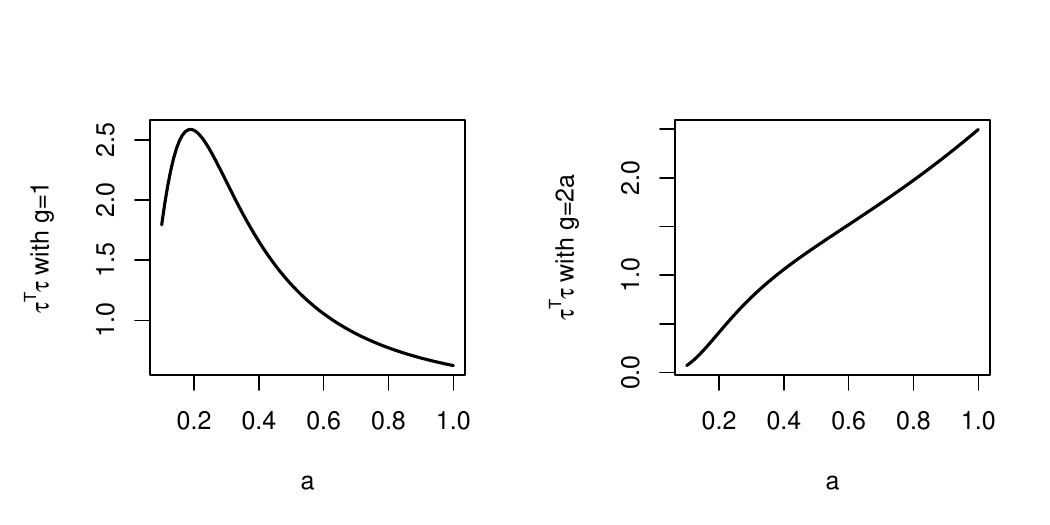}
    \caption{Trends of $\btau\Tr\btau$ as $a$ changes. The left panel uses $g=1$. The right panel uses $g=2a$. Both panels use $n=100, p=300, s=8, b=0.5, q=10$. }
    \label{Fig:tau_comparison}
\end{figure}

Fixing $p$ and $q$, $\btau\Tr\btau$ is a function of $a$ and $g$, which are of interest for controlling the amount of confounding. $g$ only appears in the numerator, so $\btau\Tr\btau$ increases with $g$ when $a$ is fixed. However, the left panel of Figure \ref{Fig:tau_comparison} shows that when $g$ is fixed, $\btau\Tr\btau$ is not a monotonic function of $a$. In the right panel of Figure \ref{Fig:tau_comparison}, it shows that after reparameterizing $g=ra$, $\btau\Tr\btau$ changes monotonically with $a$, which implies that as there are more confounding effects in the data, more bias is introduced. 

\begin{figure}[ht]
    \centering
    \includegraphics[scale=0.75]{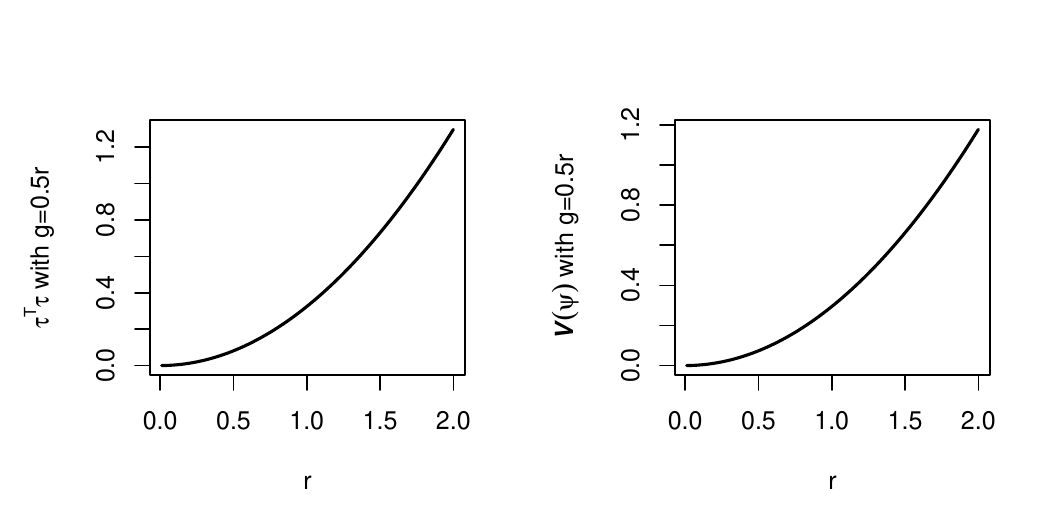}
    \caption{Trend of $\btau\Tr\btau$ as $r$ increases. $a=0.5, g=ra$. $n=100, p=300, s=8, b=0.5, q=10$. }
    \label{Fig:tau_r}
\end{figure}

$r$ controls how much confounding effect goes to the outcome relative to the features. Figure \ref{Fig:tau_r} shows that as $r$ increases, more confounding directly affects the outcome, so both $\btau\Tr\btau$ and $\Var(\psi\given\btau)$ increase. 

The parameterization of $g=ra$ makes sure that a monotonic relationship exists between $(a, r)$ and $(\btau\Tr\btau, \Var(\psi\given\btau))$, so that the monotonicity also holds for BNR, BSR, and SNR. As $a$ and/or $r$ increase, more confounding effects are added, so BNR increases and SNR decreases.

\end{appendices}

\end{document}